\documentstyle[PASJadd]{PASJ95}
\draft
\begin{document}

\title{Accretion in Gravitationally Contracting Clouds}

\author{Kohji {\sc Tomisaka}\thanks{Visiting fellow of National Astronomical Observatory.}\\[12pt]
{\it Faculty of Education, Niigata University, 
8050 Ikarashi-2, Niigata 950-21}\\
{\it E-mail: tomisaka@ed.niigata-u.ac.jp.}}

\abst{
Accretion flow in a contracting magnetized interstellar cloud was
 studied using magnetohydrodynamical simulations and a nested grid
 technique.  
First, the interstellar magnetized cloud experiences a
 ``runaway collapse'' phase, in which the central density increases
 drastically within a finite time scale. 
Finally, it enters an accretion
 phase, in which inflowing matter accretes onto a central high-density
 disk or a new-born star.  
We found that the accretion rate reaches (4 -- 40) $\times c_s^3/G$,
 where $c_s$ and $G$ represent the isothermal
 sound speed and the gravitational constant, respectively.  
This is
 much larger than the standard accretion rate of $0.975c_s^3/G$ for
 a hydrostatic isothermal spherical cloud (Shu 1977, AAA19.065.044).  
Due to the effect of an extra
 infall velocity achieved in the runaway phase ($\sim 2 c_s$),
 the accretion rate is boosted.  
This rate declines
 with time in contrast to Shu's solution, but keeps $\gtsim 2.5
 c_s^3/G$.  
The observed gas infall rate around proto-stars such as
 L1551 IRS 5 and HL Tau is also discussed.
}

\kword{Interstellar: clouds --- Accretion ---
Magnetohydrodynamics --- Stars: formation ---  Interstellar: medium}

\maketitle

\section{Introduction}

The physical processes during the course of star formation has attracted
 much attention in recent years.  
Many numerical simulations have
 been performed in order to study the dynamics of the collapse of interstellar
 clouds (Larson 1969; Penston 1969; Norman et al. 1980; Narita et al.
 1984; for a review, Boss 1987). 
Even if a cloud is assumed to be initially uniform,
 the density gradient grows in a dynamical time scale (a
 sound crossing time).  
After that, mainly the central part of the cloud contracts.  
Since the local free-fall time is proportional to $\propto (G \rho)^{-1/2}$,
 only contraction  in the central dense part proceeds.  
Thus, the central density  
 increases drastically in a finite time (within a couple of sound crossing
 times), as long as the equation of state remains isothermal.  
This is called a ``runaway collapse''.  
It is important that the runaway collapse is not stalled by the 
 centrifugal force (Norman et al. 1984), magnetic force
 (Scott, Black 1980), or even both (Basu, Mouschovias 1994).
Thus, it is concluded that if the cloud begins to collapse, the
 isothermal cloud continues its ``runaway collapse'' irrespective of
 the rotation and the magnetic field strength in the cloud.

This is a story of star formation in a cloud far from any
 hydrostatic equilibria.  
There is another model, in which the cloud
 evolves in a quasi-static fashion driven by the plasma drift
 (ambipolar diffusion; Nakano 1979, 1982, 1983, 1984; Lizano, Shu 1989;
 Tomisaka et al. 1990; Fiedler, Mouschovias 1993;
 Ciolek, Mouschovias 1994; Basu, Mouschovias 1994).  
The neutral
 molecules flow across the magnetic field lines due to the
 self-gravitational force, and the ratio of the gas density to the magnetic
 field strength in a central region increases with time.  
As a result,
 the central density increases quasi-statically.  
However, after the above-mentioned
 ratio surpasses a critical value, this quasi-static evolution no
 longer continues, and the cloud begins dynamical contraction.

Shu (1977) studied this dynamical contraction in an isothermal spherical
 cloud. 
He assumed that the contraction begins in a cloud having a
 hydrostatic density distribution with singularity at the center as $\rho(r)
 =(2\pi G r^2)^{-1}$.  Inflow begins from the center and it propagates
 outward.  
Further, the boundary between the inflowing accretion flow
 and the static, unperturbed part propagates outwardly as a
 rarefaction wave front with the sound speed.  
Therefore, this is called ``inside-out collapse.''  
Shu showed that this gives an
 accretion rate which depends only on the sound speed $c_s$ (and the
 gravitational constant $G$) as $\dot{M}_{\rm S}=0.975 c_s^3/G$, and
 is constant irrespective of time.  
However, this depends upon 
 the initial condition that the accretion begins in a
 {\em static} cloud.  
In contrast, the results of dynamical  simulations show
 that there exists a ``runaway collapse'' even before the accretion
 flow begins.  
In this Letter, we consider the complete evolution
 from the ``runaway collapse'' to the ``inside-out collapse''.  
We show that more violent accretion occurs, and that the accretion rate is
 much higher than that of Shu.  
In the next section, we describe the model.  
In section 3, numerical results are given and section 4 is devoted to
  a discussion.

\section{Model and Numerical Method}

We consider an axisymmetric hydrostatic cloud.
To initiate the cloud to collapse, we add a positive density perturbation
 to the cloud and study its evolution.

The fragmentation of a cylindrical parent cloud with axial magnetic fields
 leads to a number of clouds threaded with magnetic field
 lines and connected with each other.
We assume that one cloud is formed per length of $l_z$.
If the cylindrical cloud is long, $l_z$ is determined as the
 wavelength of the gravitational instability with the maximum growth
 rate $\lambda_{\rm MGR}$.  
The evolution in this case has been studied by
 Tomisaka (1995: Paper I, 1996: Paper II).  
These clouds also experience a runaway collapse.  
Here, we assume $l_z < \lambda_{\rm MGR}$.  When
 $l_z \ll \lambda_{\rm MGR}$, there can exist a hydrostatic configuration
 in such a cloud.
As an initial state, we adopt a hydrostatic cloud formed in the case of
 $l_z < \lambda_{\rm MGR}$.

The magnetohydrostatic equilibrium is obtained by solving the coupled
 equations (Mouschovias 1976; Tomisaka et al. 1988) of
 the Poisson equation for the gravitational potential and the
 Grad-Shafranov equation for the magnetic vector potential.
We search for solutions by an iterative method,
 the so-called self-consistent field method.
Since this solution is in hydrostatic equilibrium,
 the structure remains unchanged after time integration.
Thus, a positive density perturbation with a relative amplitude of 10\%
 is added to the central several grids. 

After choosing the units as $\rho_s$(the density on the cloud surface)
 $=c_s=4\pi G=1$, the model contains 4 parameters: the plasma beta of
 the virtual parent cylindrical cloud, $\beta$, the center-to-surface
 density ratio of this parent cloud, $F$, the length of the cloud, $l_z$, and
 the mass of the cloud, $M_{\rm cl}$.  
The assumed model parameters are given in table 1.  

As boundary conditions, we assume that far from the cloud the magnetic
 field approaches a uniform one as $(B_z, B_r)=((8\pi
 p_{\rm ext}/\beta)^{1/2},0)$, where the external pressure $p_{\rm ext}$
 is taken to be equal to the surface pressure $\rho_s c_s^2$.  On the outer
 boundary of a numerical box, the gravitational potential $\psi$
 reduces to that made by a series of point masses of $M_{\rm cl}$,
 which are separated from each other by $l_z$ on the $z$-axis.

The basic equations to be solved are the magnetohydrodynamics and 
 Poisson equations in  cylindrical symmetry.
As a numerical scheme, we employed  van Leer's (1977) monotonic
 interpolation and the constrained transport (Evans, Hawley
 1988) for the magnetic induction equation.
As a Poisson solver, we adopted
 the preconditioned conjugate gradient squared method.  
To resolve the
 fine structure near the center of the cloud, the nested grid
 technique was employed (for the detail of this numerical scheme, see
 Paper II).  
This nested grid method using 14 levels of grids has
 an effective resolution 
 corresponding to a simple finite-difference scheme using
 $10^6\times 10^6$ grid points.

\section{Results}

The initial structure of the static cloud is shown in figure 1a for L2
 (Level 2 of the nested grid system; L$n$ has twice finer resolution
 than L$n-1$).
First, the perturbation grows in proportion to $\delta
\rho \propto\exp(\omega t)$, which is characteristic to linear growth.  
Since a flow crossing the magnetic field lines is blocked
 by the field, a disk which runs perpendicular to the magnetic fields is
formed.  
After $t\sim 0.18$, the central density increases rapidly
 with time, and at $t=0.2115$ the density reaches $10^{10}\rho_s$ in
 the runaway collapse phase (the evolution in this phase was described
 previously in Papers I and II). 
Even at this time, the  global structure is
 unchanged from that shown in figure 1a.  
The structure of the central region (L7) at
 this stage is shown in figure 1b.  
Gas is contracting onto the disk
 flowing along the magnetic field lines, which run perpendicular to
 the disk, with $|v_z| \gg |v_r|$.  
Outward-facing shock waves, whose fronts are parallel to the disk,
 are formed.  
The magnetic field lines
 are pulled radially inwardly in the disk where the thermal pressure
 dominates the magnetic one. 
The cloud now enters the accretion phase.

To continue the simulation, we adopted so-called ``sink cell method,''
 in which the matter in the grids near the center is assumed to
 accrete onto the central high-density body, and to be removed from the cells
 (Boss, Black 1982).  
We assume that the sink cells distribute
 inside a distance of 16-times the grid size from the center in L7
 and only in the sink cells in which $\rho > \rho_{\rm sink}$, the excess
 density $(\rho - \rho_{\rm sink})$ and the corresponding momentum are
 both removed (here we take $\rho_{\rm sink}=10^6\rho_s$).  
The removed mass in each time steps  accounts for the increase in the mass
 of the central body.
As a result of the second assumption, sink actually occurs in a small
 number of central grids.  
Compared with a reference calculation 
 with a different size of the sink region, it is confirmed that
 the size of the sink region is not important to the result.

The structure at $\tau=0.0569$ after the beginning of accretion is
 shown in figure 1c.  
(In this Letter, $t$ and $\tau$ represent
 the time from the perturbation was given and that from the accretion began.)
It clearly shows that the magnetic fields are squeezed, and
 run almost radially near the disk.  
The shape of the magnetic field lines becomes like an hourglass. 
Comparing figures 1b and 1c, it is
 shown that the disk continues to contract, and becomes thinner during the
 accretion phase.  
>From figure 1d, it is shown that the contracting
 speed exceeds the sound speed and reaches $\simeq 5 c_s$ (near
 $r\simeq 0.02$), while in the runaway collapse phase it is not more
 than (1.5-2)$c_s$.

The accreted mass is calculated as the mass removed from the sink
 cells.  
In figure 2, we illustrate the time evolution of the accretion rate
 $\dot{M}$ normalized to $c_s^3/G$ by the thick lines.  
This clearly shows that the rate is {\em not} constant with time.  
Since the value for Shu's solution is equal to 0.975, the difference
 attains $\gtsim 40$ times just after accretion begins.  
Even in the late phase, the accretion rate is $\gtsim $2.5-times
 larger than that expected in a static cloud (0.975).
The time-averaged accretion rate ($\int_0^\tau \dot{M}\,dt/\tau$)
 is equal to $\simeq 4 c_s^3/G$. 
This is explained by the extra inflow that
the cloud achieved in the runaway phase.  
Gas within $r \ltsim 0.02$ is falling
at a speed of $|v_r| \gtsim 1.5 c_s$ (figure 1b) when accretion begins.
This enhances the accretion rate by an amount  $\sim \rho |v_r|
4 \pi R^2$ if spherical symmetry is assumed.

To see the effect of the initial condition on the accretion rate, we studied
 model B, which has a lower density contrast ($\rho_c/\rho_s$)
 and a larger mass ($M_{\rm cl}$) than those of model A.  
In this case, a density perturbation with a
 relative amplitude of 20\% is necessary for the cloud to begin
 contraction, while this is 10\% for model A.  
This means that the initial
 condition of model B is much more stable than that of model A for
 gravitational contraction.  However, model B indicates a larger accretion
 rate by contraries (thin lines of figure 2).  
It seems to result from
 the fact that the runaway collapse phase lasts for 0.96 in contrast
 to 0.21 (model A). 
This indicates that the extra inflow is considerably accelerated in 
 the runaway phase, and that the inflow achieved
 in the runaway phase plays an important role concerning the later accretion
 rate.
 
\section{Discussion}

A high accretion rate has been argued by Whitworth and Summers (1985)
 as well as Foster and Chevalier (1993) for spherical symmetric clouds.  The
 former studied various {\em self-similar} spherically symmetric
 solutions, and found that the accretion rate reaches $46.84 c_s^3/G$
 in the case that the accretion begins in the Larson(1969) and Penston
 (1969) self-similar solution at $\tau=0$ 
(The Larson and Penston
 self-similar solution itself has a constant mass inflow rate as $29
 c_s^3/G$.).  
The Larson and Penston self-similar solution
 is believed to be attained near the center of the cloud in a
 runaway (dynamical) contraction which occurs in an initially diffuse
 cloud.  
Thus, it seems natural that in the beginning of the accretion
 phase the accretion is very active as $\dot{M} \simeq$ (40--50)
 $c_s^3/G$, as shown in our MHD simulations. 
Foster and Chevalier
 (1993) confirmed this by their one-dimensional hydro-code, and
 studied cloud contraction from $\tau < 0$ to $\tau > 0$
 throughout.  
They found that the accretion rate attains $\sim 10 c_s^3/G
 \gg \dot{M}_{\rm S}$, and then gradually declines.

The accretion phase of a magnetized cloud has been studied by Galli and
 Shu (1993a, b).  
They showed that the accretion rate is not affected very
 much by the effect of magnetic fields, e.g., $\dot{M}\simeq 0.975
 c_s^3/G$;  this comes from the subtle role of magnetic fields in
 the dynamics.  
The latter fact is also confirmed in this  Letter,
 since the accretion rate is not largely different from that
 calculated for non-magnetic clouds (Whitworth, Summers 1985).
However, the accretion rate of Galli and Shu is much smaller than
 ours.  
This is explained as follows.
They considered a cloud such as a
 hydrostatic singular isothermal sphere threaded by uniform
 magnetic fields, i.e., accretion is assumed to begin in this static
 cloud.  
In contrast, since an extra infall velocity is achieved during the dynamical
 runaway collapse, our model indicates a larger accretion rate
 than does theirs.

The normalization adopted here gives the actual physical scales: for time
 $t=(4\pi G \rho_s)^{-1/2}=1.75 ({n_s}/{100\,{\rm cm}^{-3}})^{-1/2}$Myr,
 where the $n_s$ represents the particle density at the surface of the
 cloud, for an accretion rate $\dot{M}=c_s^3/G=1.62\times 10^{-6}
 (c_s/190\, {\rm m\,s}^{-1})^3$
 $M_\odot {\rm yr}^{-1}$, and for mass
 $M=\dot{M}t=c_s^3/G/(4\pi G \rho_s)^{1/2}=2.84 (c_s/190\, {\rm
 m\,s}^{-1})^3 ({n_s}/{100\,{\rm cm}^{-3}})^{-1/2}M_\odot$.  
Here, we assumed that the temperature is equal to 10 K.  
Thus, if we choose the
 above normalization, it is shown that a mass of $M=1.78~ M_\odot$ has
 accreted in $\tau=0.21$ Myr (final state of figure 2).  
The average accretion rate
 becomes $8.7\times 10^{-6} M_\odot{\rm yr}^{-1}$.  
In contrast, the model by Shu
 (1977) predicts $1.58 \times 10^{-6}M_\odot{\rm yr}^{-1}$ for
 the standard $A=2$ model and $9.04\times 10^{-6}M_\odot{\rm yr}^{-1}$ for
 the $A=4$ model.
[ This $A$ is a density parameter to specify the
 initial density distribution as $\rho(r)= A c_s^2/(4\pi G)r^{-2}$.
 The distribution coincides with that of the singular hydrostatic
 solution (Chandrasekhar 1957) only when $A=2$.  When $A > 2$, the
 solution is out of hydrostatic balance and the cloud is contracting.
 Since accretion occurs in a dynamically-contracting cloud core in our
 model, it is reasonable that the accretion rate is similar to that
 expected by the Shu's model with $A=4$.]

It is not easy to determine the accretion rate from the direct detection
 of infalling gas.  
However, there exist a number of observations.
>From a combination of the infall velocity ($0.55\,{\rm km\,s}^{-1}$) and
 the number density ($2\times 10^5{\rm cm}^{-3}$) at a radius from the
 center $20''$ of L1551 IRS 5, Fuller et al. (1995) obtained a mass
 accretion rate to this object as $\dot{M}\ltsim 1.4\times
 10^{-5}M_\odot {\rm yr}^{-1}$.  
This is rather larger than Shu's accretion rate mentioned above.  
However, this can be naturally explained by our
 model.  
Applying our models to fit this accretion rate, the age
 should be $3\times 10^4$yr (model A) or $8\times 10^4$yr (model B)
 for this object.  
This is consistent with the age determined from the
 dynamical age of the outflow, $3\times 10^4$ yr -- $10^5$ yr
 (Moriarty-Schieven, Snell 1988).  
Another example is HL Tau.  
>From the Nobeyama Millimeter Array map, Hayashi et al. (1993) estimated the
 mass infall rate at 1400~AU from the center as $5\times 10^{-6}M_\odot
 {\rm yr}^{-1}$ -- $2\times 10^{-5}M_\odot {\rm yr}^{-1}$.  
Since the accretion rate is proportional to $c_s^3$, a large accretion rate may
 be explained by a slightly higher temperature, such as 20 K.  
However,
 these accretion rates are naturally explained by our
 model without assuming a higher temperature. 
New observations with a finer spatial resolution are desired in order
 to resolve the inflow velocity profile and to determine the dynamics in the
 contracting cloud further.
\par
\vspace{1pc} \par
The work was supported in part by Grant-in-Aids for
 Scientific Research from Ministry of Education, Science, Sports and
 Culture (07640351, 07304025).

%\clearpage
\section*{References}
{\small
\re
  Basu S., Mouschovias T. Ch. 1994, ApJ 432, 720
\re
 Boss A. P. 1987, in Interstellar Processes (Reidel: Dordrecht), ed
 D. J. Hollenbach, H. A. Thronson, Jr. p.321
\re
 Boss A. P., Black D. C. 1982, ApJ 258, 270
\re
 Chandrasekhar S. 1939, An Introduction to the Study of Stellar Structure
 (University Chicago Press: Chicago), section 22
\re
 Ciolek G. E., Mouschovias T. Ch. 1994, ApJ 425, 142
\re
 Evans C. R., Hawley J. F. 1988, ApJ 332, 659
\re
 Fiedler R. A., Mouschovias T. Ch. 1993, ApJ 415, 680
\re
 Foster P. N., Chevalier R. A. 1993, ApJ 416, 303 
\re
 Fuller G. A., Ladd E. F., Padman R., Myers P. C.,  Adams F. C. 1995,
 ApJ 454, 862
\re
 Galli D., Shu F. H. 1993a, ApJ 417, 220
\re
 Galli D., Shu F. H.  1993b, ApJ 417, 243
\re
 Hayashi M., Ohashi N., Miyama S. M. 1993, ApJ 418, L71
\re
 Larson R. B. 1969, MNRAS 145, 271
\re
 Lizano S., Shu F. H. 1989, ApJ 342, 834
\re
 Moriarty-Schieven G. H.,  Snell R. L. 1988, ApJ 332, 364
\re
 Mouschovias T. Ch. 1976, ApJ 206, 753
\re
  Nakano T. 1979, PASJ 31, 697
\re
  Nakano T. 1982, PASJ 34, 337
\re
  Nakano T. 1983, PASJ 35, 87
\re
  Nakano T. 1984, Fund. Cosmic Phys. 9, 139
\re
 Narita S., Hayashi C., Miyama S. M. 1984, Prog. Theor. Phys.
 72, 1118
\re
 Norman M. L., Wilson J. R.,  Barton R. T. 1980, ApJ 239, 968
\re
 Penston M. V. 1969, MNRAS 144, 425
\re
  Scott E. H., Black D. C. 1980, ApJ 239, 166
\re
 Shu F. H. 1977, ApJ 214, 488
\re
 Tomisaka K. 1995, ApJ, 438, 226 (Paper I)
\re
 Tomisaka K. 1996, PASJ, 48, No.5 (Paper II)
\re
 Tomisaka K., Ikeuchi S., Nakamura T. 1988, ApJ 326, 208
\re
 Tomisaka K., Ikeuchi S., Nakamura T. 1990, ApJ 362, 202
\re
 van Leer B. 1977, J. Comp. Phys. 23, 276
\re
 Whitworth A.,  Summers D. 1985, MNRAS 214, 1 
}

\clearpage
\begin{table*}
\begin{center}
Table~1.\hspace{4pt}Model parameters.\\
\end{center}
\begin{tabular}{cp{2cm}p{2cm}p{2.5cm}p{2.5cm}p{2.5cm}p{2.5cm}}
\hline
\hline
 Model &
 $\beta^\ast$ &
 $F^\dagger$ &
 ${\rho_c/\rho_s}^\ddagger$ &
 $M_{\rm cl}^{\S}$ &
 ${l_z}^{\P}$  \\
\hline
A...... &  1  & 100 & 1000 & 37.96 & 12.73\\
B...... &  1  & 100 & 100 &  46.4 & 12.73\\
\hline
\end{tabular}
\vspace{6pt}\par\noindent
$*${The plasma $\beta$ in the parent cylindrical cloud.}
\par\noindent
$\dagger${The center-to-surface density ratio in the parent cloud.}
\par\noindent
$\ddag${The center-to-surface density ratio of the hydrostatic cloud.}
\par\noindent
$\S${The mass of the cloud.}
\par\noindent
$\P${The length of the parent cloud.}

\end{table*}

\clearpage
\centerline{Figure Captions}
\bigskip
%%% fig.1 %%%
\begin{fv}{1}{1cm}
{
{(a)} Initial state ($t=0$). 
Density contour lines (closed lines) and
 magnetic field lines (running horizontally) are plotted for the L2 grid.
The steps of the density contours are taken as 
 $\rho=10^{3n/20}$ ($n=0,1,\ldots 20$).
 The $z$-axis, the symmetry axis, is taken 
 horizontally.
{(b)} The structure at the end of the runaway collapse phase 
($t=0.2115$) in the L7 grid.
 A disk is formed running perpendicular to the global magnetic field lines.
 The size of the grid is $2^{5}=32 $ times finer than that of (a).
{(c)} The structure in the accretion phase
  at $t=0.2684$ ($\tau=0.0569$). 
The size of the grid is the same as (b).
{(d)} Radial distributions of the density ($\rho$: solid line),
 the radial velocity ($v_r$: dotted line) and
 the magnetic field strength ($B_z$: dashed line) on the equatorial
 plane ($z=0$).
}
\end{fv}
%%% fig.2 %%%
\begin{fv}{2}{1cm}
{
Mass accretion rate  against the age after the core formation
 $\tau$.
The rate is normalized to $c_s^3/G$.
The solid and dotted lines represent the instantaneous accretion rate
 as $\dot{M}(\tau)$ and the time averaged one as $\int_0^\tau\dot{M}(t)dt/\tau$, respectively. 
The thick and thin lines correspond to models A and B, respectively. 
The value obtained by Shu (1977) is 0.975.
On the upper and right axes, typical {dimensional} values are plotted 
 with assuming the conversion factors in section 4 as 
 $t=(4\pi G \rho_s)^{-1/2}=1.75 ({n_s}/{100\,{\rm cm}^{-3}})^{-1/2}$Myr and
 $\dot{M}=c_s^3/G=1.62\times 10^{-6} (c_s/190\, {\rm m\,s}^{-1})^3M_\odot 
{\rm yr}^{-1}$.
}
\end{fv}
\clearpage
%\clearpage
%\begin{figure}
%\noindent\hspace*{-3cm}{\it (a)\hspace{7.5cm}(b)}\\
%\plottwo{fig1a.eps}{fig1b.eps}

%\noindent\hspace*{-6.7cm}{\it (c)\hspace{7.5cm}(d)}\\
%\plottwo{fig1c.eps}{fig1d.eps}
%\caption{}
%\end{figure}

%\begin{figure}
%\plotonec{fig2.eps}
%\caption{
%}
%\end{figure}
\end{document}